# Attractors in residual interactions explain the differentially-conserved stability of Immunoglobulins

Anirban Banerji
Centre for Molecular Modeling, Indian Institute of Chemical Technology, Hyderabad-500607, India
anirbanab@gmail.com

**Abstract**
Proteins belonging to immunoglobulin superfamily(IgSF) show remarkably conserved nature both in their folded structure and in their folding process, but they neither originate from very similar sequences nor demonstrate functional conservation. Treating proteins as fractal objects, without studying spatial conservation in positioning of particular residues in IgSF, this work probed the roots structural invariance of immunoglobulins(Ig). Symmetry in distribution of mass, hydrophobicity, polarizability recorded very similar extents in Ig and in structurally-closest non-Ig structures. They registered similar symmetries in dipole-dipole, $\pi$-$\pi$, cation-$\pi$ cloud interactions and also in distribution of active chiral centers, charged residues and hydrophobic residues. But in contrast to non-Ig proteins, extents of residual interaction symmetries in Ig.s of largely varying sizes are found to converge to exactly same magnitude of correlation dimension - these are named 'structural attractors', who's weightages depend on ensuring exact convergence of pairwise-interaction symmetries to attractor magnitude. Small basin of attraction for Ig attractors explained the strict and consistent quality control in ensuring stability and functionality of IgSF proteins. Low dependency of attractor weightage on attractor magnitude demonstrated that residual-interaction symmetry with less pervasive nature can also be crucial in ensuring Ig stability.

**1. Introduction:** The immunoglobulin superfamily(IgSF) is a heterogeneous group of proteins built on the common structural platform, the 'Immunoglobulin-like beta-sandwich'(Ig$\beta$SW) fold. Studying IgSF assumes importance because immunoglobulins(Ig) play key roles in defense mechanism of an organism against pathogens. The Ig$\beta$SW fold (Amzel and Poljak, 1979) can easily be detected as one of the most prevalent folds (Gerstein and Levitt, 1997; Wright et al., 2004) encoded by the human genome; more than 750 genes in human genome have been found to encode proteins with at least one Ig. fold (Berg et al., 2006). Not surprisingly, such enormous degree of many-to-one mappings have resulted in many sequences folding into an "average core structure" (Gerstein and Altman, 1995; Clarke et al., 1999); whereby, despite having significant variation in sequences, members of IgSF are found to demonstrate similar overall structural commonality (Bork et al., 1994; Parker et al., 1998). Moreover, folding pathways of IgSF proteins are found to share common characteristics (Clarke et al., 1999).

Remarkably, regardless of having these commonalities, proteins belonging to IgSF are found to have little or no detectable evolutionary or functional relationship among themselves (Halaby and Mornon, 1998; Clarke et al., 1999). Indeed, the observation that although IgSF proteins are clustered into subfamilies based on sequence similarity, different subfamilies show no significant sequence identity (Lappalainen et al., 2008) – adds to the complexity of the problem. – Which (deep) principle of protein structure organization, then, can explain the aforementioned consistency in precision of the many-to-one sequence-to-structure mapping scheme? More categorically, what unique character of Ig structures ensure that the diverse array of sequences follow such extensive many-to-one sequence-to-structure mapping scheme? Posed alternatively, how different are the structure organization principles in Ig proteins from that in proteins belonging to other fold families?

We start answering by observing that, though the spectrum of Ig.s can be subdivided into superfamilies with no detectable evolutionary or functional relationship, folding pathways of these proteins are conserved across superfamilies (Clarke et al., 1999; Wright et al., 2004). This implies that the canonical set of interactions constituting the folding nucleus of any of the Ig families, remain conserved; although the details of sidechain interactions may vary from case to case (Abkevich et al., 1994; Shakhnovich et al., 1996; Shakhnovich, 1997; Ptitsyn, 1998; Mirny et al., 1998; Ptitsyn and Ting, 1999). Such a process of formation of conserved structural core, especially from sequences sharing no detectable evolutionary relationship (Gerstein and Altman, 1995;

Clarke et al., 1999) implies that stability of Ig structures is more dependent upon the nature and spatial extent of interactions than on positioning of particular residues, since the later possibility would have made it obligatory for the sequences to be highly similar. To quantify the *general nature* of residual (and in general, biophysical) interactions, one may resort to calculation of symmetry embodied by each of these interactions (Dewey, 1997). Thus, to decipher the differentially-conserved root of Ig stability, we quantified the symmetry in distributions of each of the biophysical interactions that are responsible to ensure the Ig "core-formation". The magnitude of symmetry thus quantified, upon being detected to remain invariant across each of IgSF's many-to-one sequence-to-structure mapping schemes, may provide a possible way to identify the core-forming canonically important interactions. But then, existence of distinct superfamilies in IgSF suggests that the extents of characteristic symmetry in distributions of biophysical properties in IgSF domains vary subtly from one case to another. Hence we attempt to decipher the nested patterns of differentially conserved symmetry in distributions of various biophysical properties in IgSF, before quantifying the subtle differences in the symmetry of residual and biophysical interactions in general terms.

Why quantify symmetry of interactions? This question entails two small questions, first, why study interactions (rather than studying spatial conservation of (key) residues)? Second, why study interactional symmetry (rather than studying particular functionality of certain (key) interactions)? Indeed one notes that studies over the last two decades have attempted to trace the root of Ig stability and functionality by either identifying the conserved nature of placement of particular residues to particular spatial location of Ig proteins (Gerstein and Altman, 1995; Improta et al., 1996; Mirny and Shakhnovich, 1999; Ptitsyn and Ting, 1999; Halaby et al., 1999; Reddy et al., 2001 (in somewhat broader scope); Cota et al., 2001; Honegger et al., 2009); or, by investigating the conservation of common structural features of residual arrangement with tools derived from Euclidean geometry (Gelfand et al., 1998; Stoyanov et al., 2000; Silverman, 2007); or, by comparative analysis of topologies (Overington et al., 1990); or by some superposing one of these philosophies on any other (for example, Honegger and Plückthun, 2001; Wright et al., 2004(and some references therein)). Many of these had attempted to explain Ig-structural invariants by studying the (conserved) spatial arrangement of relative-positioning of some or the other Ig-residues.

Informative as these approaches undoubtedly were, the general principles that ensure differential-conservation of dual phenomena of Ig stability and functionality – could not be deciphered by them. Geometric regularity in relative-spatial positioning of residues may indeed serve as a quick handle to assess similarity of structural core ; but then one notes that a) many residues share similar biophysical/biochemical characteristics, b) many of the interactions are not merely local and individual in acting between two particular residues, but are cooperative across the entire space of protein (Batey and Clarke, 2006; Lindberg and Oliveberg, 2007) and c) strength of biophysical/biochemical interactions are inherently context-dependent (Rost, 1997; Zarrine-Afsar et al., 2005); whereby, one may hypothesize that it is the conserved character of biophysical *interactions* among these residues that assumes more fundamental nature than the spatial positioning of residues themselves. But then, to understand the problem from a general perspective, one requires connecting the separate local interactional-schemes among particular residues, to the global level of interactional-invariance for the emergence of any biophysical property. This can be achieved by quantifying the *symmetry* in distributions of each of the biophysical properties under consideration. To quantify the subtle differences that characterize stability profile of each subfamily of Ig proteins, the *extent of symmetries* in biophysical interactions prevailing among proteins belonging to any particular superfamily of IgSF can be contrasted to that prevailing among proteins belonging to some other superfamily of IgSF. To answer the other question, viz., how and to what degree does the stability profile of an Ig protein differ from that of a non-Ig protein, the extent of symmetries in biophysical interactions prevailing in a Ig protein can be contrasted to that prevailing in *structurally most-*

*similar* non-Ig protein. Thus, banking on an unified scheme to describe protein stability from interactional symmetry, both the questions, viz.,

> **i:** how different are the principles of structural organization in Ig from that in others? (posed alternatively, how similar are the principles of structural organization in Ig than what prevail in other types of (comparable) protein structures?)
>
> and
>
> **ii:** how different are the principles of structural organization in one superfamily of IgSF from that in other superfamilies of IgSF? (posed alternatively, how similar are two superfamilies of Igs with respect to any particular aspect of structural organization?)

- could be approached.

**2.1. Materials:** Dataset used for studying interactional symmetry in different classes of Ig proteins is taken from (Halaby et al., 1999). In order to quantify the difference in the extent of such symmetry existing in Ig and non-Ig proteins, each Ig molecule of the aforementioned dataset was subjected to structural comparison against the whole universe of protein structures using DALI(Holm and Rosenström, 2010); whereby, for every Ig protein in the dataset, the non-Ig protein that matched its structure to the best possible degree – was identified. To maximize the number of Ig-vs-non-Ig pairs, a non-strict resolution (≤ 3.5Å) was considered for non-Ig structures. More importantly, it was made sure that none of the proteins considered contained any "disordered" regions (as defined in (Dunker et al., 2001)), because presence of those would have severely influenced the symmetry calculations at all levels. However, lenient as they are, upon imposing these selecting criteria, 32 Ig and 15 non-Ig proteins were found suitable for the top-down (viz., MFD, HFD, PFD, macroscopic CD calculations of different types) analyses; for studying 210 residue-residue interaction symmetries, 32 Ig and 13 non-Ig proteins were found suitable.

**2.2. Methods:** It was established by previous studies (Enright and Leitner, 2005; Banerji and Ghosh, 2011) that mass distribution across protein space follows a fractal scaling. Extending the idea of residue-specific 'atomic hydrophobicity' (Kuhn et al., 1995) a previous study (Banerji and Ghosh, 2009) had shown that, just like the mass distribution characteristics, hydrophobicity distribution within proteins also follow fractal scaling. Each protein residue is characterized by its intrinsic magnitude of polarizability (Song, 2002). Though interior electrostatics of proteins is anisotropic, distribution of residual polarizability within protein space is also found to follow a fractal scaling (Banerji and Ghosh, 2009). Extents of mass, hydrophobicity and polarizability fractal dimensions(FD) were calculated from respective scaling relationships: $M\ R^{MFD}, H\ R^{HFD}, P\ R^{PFD}$; where M, H and P represent the global radial distributions of mass, hydrophobicity and polarizability filling the entire space of an arbitrarily chosen protein; MFD, HFD and PFD represent mass hydrophobicity and polarizability FD; R represents the characteristic length-scale, that, when measured from center of mass (viz., or any other suitable stability center) of the protein, can detect the scale-free nature in the distributions of mass, hydrophobicity and polarizability in protein.

Interactional symmetry was measured with 'correlation dimension'(CD) (Grassberger and Procaccia, 1983; Takens, 1985). While FD treated protein atoms and residues as individual entities, CD quantified the symmetry of interactions among individual residues and separately, among classes of residues. Generalizing the idea of CD, a previous work (Banrji and Ghosh, 2011) had demonstrated that the symmetry in interactions among biophysical entities (say, among peptide-dipoles, among [π-cloud]–[π-cloud], among cation–[π-cloud], etc.) could be calculated too. The 'correlation function' $C_N(\varepsilon)$ is defined as:

$$C_N(\varepsilon) = \frac{2}{N(N-1)} \sum_{0 \leq i < j \leq N} H\left(\|C_i^\beta - C_j^\beta\| < \varepsilon\right) \qquad \text{Eq}^n\text{ - 1}$$

where H(X) is the Heaviside function whose value is 1 if the condition X is satisfied, 0 otherwise; and $\|.\|$ denotes the distance induced by the selected norm within the structural space of folded protein. The sum $\sum_i H\left(\|C_i^\beta - C_j^\beta\| < \varepsilon\right)$ is the number of residues (represented by the coordinates of their respective $C^\beta$ atoms ($C^\alpha$ for GLY)) within the distance $\varepsilon$ of each other (numerically, $C_i^\beta - C_j^\beta \leq 8\text{Å}$). The CD can then be defined as:

$$CD = lim_{\varepsilon \to 0} \, lim_{N \to \infty} \frac{log C_N(\varepsilon)}{log(\varepsilon)} \qquad \text{Eq}^n - 2$$

Amongst 20 residues of a protein, there can be 210 [=(190($^{20}_2 C$)+20)] types of possible residue-residue interactions. Thus 210 CD magnitudes were calculated for each (Ig, non-Ig alike) protein considered. These magnitudes were averaged for all the non-redundant proteins in respective SCOP folds; which in turn, were averaged to the level of SCOP classes. Conceptually, CD can be considered as a generalization of both contact matrix framework (Plaxco et al., 1998; Di Paola et al., 2013) and the gamut of multifarious frameworks to study residual interactions (to name a few: Narayana and Argos, 1984; Miyazawa and Jernigan, 1996; Bahar and Jernigan, 1997; Samudrala and Moult, 1998; Betancourt and Thirumalai, 1999; Mayewski, 2005; Feng et al., 2007; Bolser et al., 2008; Little and Chen, 2009; Tress and Valencia, 2010; Di Lena et al., 2012; Eickholt and Cheng, 2012). CD is different from these because it attempts to quantify the symmetry in spatial distribution of each of residual interactions, whereby the role of each of the pairwise interactions can be connected to the global organizational principles of protein quaternary structures, which themselves are changing as a function of evolutionary pressure.

Any particular [$RES_i$-$RES_j$] CD magnitude simply quantifies how pervasively symmetrical the [$RES_i$-$RES_j$] proximity is across the protein space; in other words, how "space-filling" is the effect of the symmetry of [$RES_i$-$RES_j$] distribution - within a protein. Thus, if it is found that in some arbitrarily chosen protein named ABC, the $CD_{P-Q}$ = 2.223 and $CD_{X-Y}$ = 2.387, we can infer that in protein ABC, the symmetry of interaction (viz. symmetry in the spatial distribution of proximity) between residue X and residue Y is more, than that observed between residue P and residue Q. Furthermore, the effect of self-similar symmetry in interaction between residue X and residue Y is 0.164 (2.387-2.223) space-filling unit more than that found to exist for residue P-residue Q interaction, in protein ABC.

Though we note that a previous work (Abraham et al., 1986) had outlined the theoretical framework to calculate dimension of attractors from a limited datasets, for the context of the present problem this framework is found to be idealistic. To establish the relevance of theoretical framework in the paradigm of protein structures, we note that structural attractors(SAs) can only be identified for a set of proteins and not for individual proteins. The SAs are certain preferred magnitudes (extents) of symmetry of residual interactions to which the other possible magnitudes (extents) of symmetry, converge upon. While convergence of interaction symmetries to certain preferred symmetry-magnitude forms the second step, the first step to identify SA involves construction of a residual-interaction phase plane, constituted by **a:** sequence length of each of the proteins that these residues are part of, **b:** magnitude of symmetry of their interactions in each of the proteins they are part of. Strength of such a SA is calculated as:

Attractor weightage =
[(Number of interactional symmetries with exactly same magnitude)*((maximum – minimum) sequence length for which the interactional symmetry converged upon the magnitude concerned)] * [Population in the basin of the attraction, for the SA in question, normalized by the total number of interactional symmetry magnitudes considered].
  The first component of this formula quantifies the depth of a SA, it quantifies the innate tendency of the interactional symmetries to match the attractor magnitude exactly (to second decimal place). Protein properties, most frequently, are known to be functions of protein length (viz., the number of residues). But the beauty of FD and CD is that they, ideally, quantify the scale-free symmetry of the systemic properties; whereby, despite having significant differences in length, interactional symmetries between any two residues A and B in any related set of proteins can have exactly the same space-filling effect. But then, converging to the second decimal place of the magnitude of interactional symmetry in spite of having differences in sequence length (often in the order of > 100) is a non-trivial occurrence. To underline the significance of this non-serendipitous convergence for systems (proteins) of hugely different sizes, the attractor population is multiplied with difference in protein length.

Second part of the formula, viz. basin of the attractor, quantifies neighborhood behavior of the SA; it counts the number of cases that show tendency to converge onto the exact magnitude of SA. Normalized count of symmetry magnitudes in basin of attraction was calculated to account for the difference in number of Ig and non-Ig proteins. Basin of an attractor was calculated as:
[((normalized)basin of (+0.25) space-filling unit) + ((normalized)basin of (-0.25) space-filling unit)] ; where, we calculate the (normalized) basin of attractor for any SA as:
[number of points in ±0.25 CD magnitude in Res[A]-Res[B] interaction) / total number of proteins considered]

Choice of the space-filling unit 0.25 is a result of trade-off between:
**a)** very strict convergence criterion (resulting into extremely narrow basin, which would have excluded many possible symmetries with comparable extent of magnitude to that of the attractor)
and
**b)** broad convergence criterion (which would have (wrongly) identified largely deviating extents of interactional symmetry magnitudes as a part of the basin of a SA, and in turn, would have failed to decipher the specific principles that ensure the differentially-conserved character of Ig proteins.

**3. Results and Discussions:**
*3.1: General patterns in Ig and non-Ig structural organization:* Upon being sorted with respect to their attractor weightages, 25 strongest SAs for Ig proteins are enlisted in **Table-1**; SAs for their structurally most similar non-Ig analogues are given in **Table-2**. The fact that SAs can force residue-residue interaction symmetries to converge to some particular magnitude even if the residue pair is part of some other protein with significantly disparate number of monomers, merely underlines the role of SAs as a stability determinant for proteins. **Fig.-1** and **Fig.-2** demonstrate the extent of aforementioned convergence of residual interaction symmetries to extremely selective range of CD magnitudes, which thereby constitute a SA. The degree of these convergences, alongside the interplay of parameters influencing the attractor weightage, is detailed in **Table-1** and **Table-2** (and also in **Supplementary Material-1** in complete details).

Though, the considered non-Ig proteins were the closest matches to Ig structures from the entire of universe of known protein structures and though, the sizes of non-Ig protein were found to be larger than that of the Ig proteins in general, the maximum attractor weightage of the Ig-SAs are found to be (almost) double than that found for non-Ig-SAs. Also, the attractor weightages for residue-residue interaction symmetry for the non-Ig proteins are found to decay much faster than that observed for the Ig-proteins (**Fig.-3**), this is demonstrated by the fact that for any arbitrarily-ranked Ig-SA, the Ig-SA weightage is found to be more (roughly twice) than that achieved by non-Ig-SAs, throughout the entire range of attractor weightage. The slow decay of Ig attractor weightage profile suggests that the key-factor that distinguishes Ig structural stability is in the way nature assigned weightages (viz. importance) to Ig-SAs across the ranks. The inherent differences of structural organization schemes between Ig and non-Ig proteins are discussed below.

Surprisingly, though the extents of residual interaction symmetries in Ig proteins show marked convergence to certain precise magnitudes (in contrast to the lack of it in their structurally-most-similar non-Ig counterparts), symmetry of interaction among other structural entities in Ig proteins recorded (more or less) similar behavior as recorded in the non-Ig proteins. Hence, as **Fig.-5** demonstrates, the interaction symmetry among peptide dipoles in Ig proteins is found to register similar characteristics as that found in non-Ig proteins. Similarly, symmetry in distribution of [π-cloud]-[π-cloud] interactions in protein space, symmetry in cation-[π-cloud] interactions were found to record similar extents in Ig and non-Ig proteins. At the level of residue- classes, the symmetry of spatial distribution of charged residues, hydrophobic residues, active chiral centers – all registered similar patterns in Ig and non-Ig proteins. Likewise, from a top-down perspective, the Mass-Fractal-Dimension(MFD), the Hydrophobicity-Fractal-Dimension(HFD), the Polarizability-Fractal-Dimension(PFD) in Ig and non-Ig proteins – all recorded similar behavior.

Thus, it is only at the level of residue-residue interaction symmetries that the difference in organization between Ig and (structurally-most-similar) non-Ig proteins could be detected. Parameters that either attempt to characterize structural organization at the level of entire protein (like MFD, HFD, PFD), or, parameters defined at various other levels of structural organization (like dipole-dipole interactions, π-π interactions, etc.) failed to decipher the archetypal symmetry signatures that separate Ig from non-Ig proteins.

***3.2: Which influences the strength of SA more, is it the ability to ensure convergence across protein sizes, or, is it the superior population in the basin of attraction for the attractor? What are the biological implications of these?*** SAs represent specific magnitudes of residue-interactional symmetry that act as consistent determinants of protein stability, for a set of proteins sharing common scaffold. However, since proteins of diverse sizes can populate any single structural scaffold and since there cannot be a consistently perfect one-to-one mapping between spatial position in a scaffold and position a particular amino acid, it becomes a highly improbable event for residual interaction symmetries to assume exactly the same magnitude across proteins of different sizes. As the present study shows, though highly improbable, such occurrences are observed in protein structures; where the magnitude that the residual interaction symmetries converge to, are defined as SAs.

But then, although it is necessary for the proteins to conform to specific magnitudes of residual interaction symmetries, due to complex interplay of numerous structural/topological and evolutionary constraints, not many proteins can ensure that the magnitude of a particular residual symmetry will match exactly (viz., to the second decimal place) to that of the SA's. Hence, in many cases, the recorded magnitudes of residual interaction symmetries are found to differ from that of their respective SA's ; frequently, only by a single decimal place. These magnitudes, though not "exactly" matching with that of the SA's, represent nevertheless the inherently strong tendency of the proteins to ensure that the residual interaction symmetry in question, match the specific precise extent of residual interaction symmetry, defined by the magnitude of SA. – These closely (but not exactly) matching residual interaction symmetry magnitudes constitute the basin of attraction for the SA in question. Hence, the ability of any SA to ensure convergence of residual interaction symmetry magnitudes to an exact magnitude and the population of residual symmetries in the basin of attraction for the SA in question, are not two antagonistic features; it is just that the prior represents the extent of precision to which the residual symmetries should converge to, whereas the later quantifies the permitted allowance around the SA magnitude.

The formula constructed for calculation of attractor weightage for a SA assigns equal weightage to,
a: the number of residual interaction symmetries sharing an exact magnitude of CD,
b: the difference of protein sizes that these exactly matched CDs are derived from,
and
c: the population in basin of attraction for the SA (normalized by the total number of elements in the set (viz., the total number of proteins)).

The formula had to take into account the contribution from each of these three parameters in order to ensure comprehensiveness in the quantification of attractor behavior. Equal weightage were assigned to each of these three factors, because the relative importance of these in determining a SA may change for one structural scaffold from another, in general.

For the Ig proteins, attractor weightage is found to be strongly dependent upon the strength of the SA to ensure the convergence of residual-interaction symmetries (measured with CD) to a specific magnitude (viz. that of the SA), though the protein sizes that these residual-interaction CDs are derived from, may differ by a large margin. Correlation coefficient between attractor weightage and convergent sequence length for Ig proteins recorded a super-high dependency, 0.886. However, the other component, basins of attraction for SAs in Ig proteins, are not found to influence respective attractor weightages significantly; whereby, the correlation coefficient between attractor weightages and population in the basin of attraction registered a paltry 0.227, one-fourth of that observed between attractor weightages and length difference in convergent sequences. Such sharp difference between these two dependencies for Ig proteins, unambiguously points to a strict scheme of quality control; which, to ensure Ig-stability, permits only exact match of magnitudes of residual interaction CDs and does not allow the other CD magnitudes (even though, differing by merely 1 decimal point to that of the SA) to influence the convergence capability of SAs.

Interestingly, in stark contrast to the behavior of Ig stability determinants, it is found that populations in basins of attraction of Non-Ig SAs have a significant influence on determining attractor weightages. Correlation coefficient between them registered a dependency ~0.628, significantly higher than that observed for Ig proteins (viz., 0.227). Correlation coefficient between attractor weightages and differences in protein sizes that the structurally mosr-similar non-Ig SA can manage to converge to itself, is recorded to be 0.929 – which is slightly higher than that observed for the Ig proteins (viz., 0.886).

Hence, the crucially differentiating factor that separates the Ig-stability scheme from non-Ig stability schemes is found to be rooted in their handling of populations in basins of attraction for respective SAs. Activities of immune system inevitably demand zero fault tolerance. Thus the building blocks of immune systems, Ig proteins, need to necessarily enforce the strict condition that any of its pairwise interaction symmetry magnitude must be matching to the respective SA magnitude, exactly. This absolute requirement of matching the SA magnitude is found to result in a Booleanesque paradigm, where an Ig-pairwise-interaction-CD either matches the SA magnitude to the second decimal place or doesn't populate the basin at all. – This behavior accounts for the low (CC=0.227) correlation coefficient between attractor weightage and basin population. However, for the non-Ig proteins, even though they are structurally most-similar to Ig proteins, nature did not find it necessary to enforce the Ig-like absolute strictness in pairwise interaction symmetry's matching their SA magnitude. Thus for non-Ig proteins, the pairwise interaction CD magnitudes that closely resembled the respective SA values (without matching it exactly) are found to contribute non-trivially to attractor weightage, registering the correlation-coefficient ~ 0.63. It showed that acceptability of such non-exact pairwise interaction symmetries is permitted for non-Ig proteins, in sharp contrast to strict quality control observed for Ig protein organization.

Although no SA could be found in the symmetry profile of non-Ig VAL-ILE or PRO-ASN or GLU-LEU interactions (**Fig.-2**), (almost) horizontal stretches for a small range of ordinate could be spotted in them too. This merely suggests that not only the Ig structures, but proteins belonging to other conserved folds may also be characterized by the typical SA (and basin behavior) of their own.

To what extent does the attractor magnitude influence attractor weightage in Ig and non-Ig proteins is discussed in **Supplementary Material-2**.

*3.3: A note on chemical nature of structural attractors in Ig and non-Ig proteins.* No clear "easy pattern" could be spotted in chemical character of residual interactions in **Table-1** and **Table-2**. Nonetheless, upon comparing these tables one observes a larger presence of hydrophobic interaction symmetries in the list of 25 most weighty Ig SAs. Importance of hydrophobic residues in ensuring Ig stability has been reported previously too, albeit from a 'particular residue in particular protein'-centric perspective. The present work validated these assertions from a completely general and alternative perspective. The commonalities and differences in the (percentage) populations of individual residues in 25 Ig and non-Ig SAs with highest attractor weightages are depicted in **Fig.-4**.

The absence of HIS and MET from both Ig and non-Ig populations certainly strikes as a notable commonality across the populations. Equally striking is the absence of any strong pattern among the set of residues with ≥ 8% presence in either population (viz., ALA, VAL(only in non-Ig), LEU(only in Ig), ILE, THR, LYS, GLU). In this context we note that CYS, TRP and GLY, in either of population have registered a low population. This may appear to be surprising, given that TRP and CYS are two most conserved residues. However, it is easy to note that the sheer importance of TRP and CYS arise from their specific particular interactions and not from the overall symmetry profile of their interactions, which is precisely what is quantified by the CD. This is why, apparently unremarkable residual interaction symmetries (viz., that of THR-LYS, GLU-LEU, etc. in Ig and ALA-GLN, LYS-VAL, etc. in non-Ig populations) find their places among the SAs with highest attractor weightages.

*3.4: Why other symmetry measures failed to detect the unique organization principles of Ig proteins but residual-interaction could?* Measures of (non-crystallographic) protein symmetry are many. MFD, HFD, PFD could quantify the fractal symmetry in mass, hydrophobicity and polarizability distribution by studying proteins in their entirety, viz., from a top-down perspective (**Fig.-5.1, 5.2, 5.4**). CD among active chiral centers (viz., all the $C^\alpha$ atoms and $C^\beta$ of THR and ILE), CD among charged residues and CD among hydrophobic residues could quantify the symmetry in the spatial arrangement of the proximity of residues; so did $CD^{cation-\pi}$ and $CD^{\pi-\pi}$ among residues with π-cloud and cationic side-chains (**Fig.-5.7, 5.8**). Since IgβSW is an extremely conserved fold, finding a non-Ig protein belonging to a fold different from IgβSW fold and yet registering a statistically-significant structural match with a protein belonging to IgβSW fold, was difficult. Due to the difference in number of considered structures (32 Ig versus 15 non-Ig), **Fig.-5** plots do not look exactly similar; however, trends in the plots indicate the similarity clearly.

Though the role of hydrophobicity in Ig.s is well-documented in particular cases (to mention a few, Clarke et al., 1999; Hagihara et al., 2007; Honegger et al., 2009); on a general scale, the extent of symmetry in distributions of hydrophobicity (**Fig.-5.2.B**) and mass (**Fig.-5.1.B**) across Ig proteins of varying lengths are found to register similar trend as those observed in non-Ig proteins (viz. (**Fig.-5.2.A**) and (**Fig.-5.1.A**), respectively). Confirming these trends, the symmetry of interactions among hydrophobic residues in Ig (**Fig.-5.10**) is found to register very similar profile across proteins of different lengths as observed in case of structurally most-similar non-Ig proteins. Moreover, the amount of unused hydrophobicity (= HFD - MFD (Banerji and Ghosh, 2009)) in Ig of differing lengths are found to be (almost) identical to their magnitudes measured in non-Ig proteins (**Fig.-5.3**). However, even after considering the disparity in number of structures considered, the interaction symmetry among peptide dipoles (**Fig.-5.5**) and that among HIS, ARG, LYS, GLU and ASP, viz. the charged residues (**Fig.-5.6**) demonstrated that the extents of these symmetries in Ig proteins fall behind (by a small margin) in comparison to those measured for non-Ig proteins. Surprisingly however, with respect to the distribution of residual-polarizability (**Fig.-5.4**), as also with respect to symmetry of [π-cloud]-[π-cloud] (**Fig.-5.7**) and cation-[π-cloud] interactions (**Fig.-5.8**), Ig and non-Ig proteins registered very similar behavior. Interestingly, though matching in their overall profile, the symmetry in distribution of active chiral centers in Ig proteins is found out to be a fraction less than what is measured in their structurally most-similar non-Ig proteins (**Fig.-5.9**). But none of these could detect the prominent differences in Ig organization from non-Ig organization, something that the residual interaction symmetry could. Why?

Though MFD, HFD and PFD could quantify respective fractal symmetries, these were calculated by considering the entire set of individual residues. CD calculations, in contrast could quantify the interactional symmetry among the respective type of residues (say, the hydrophobic residues, positively-charged residues, etc.) as demanded by the present problem. But such classification involves coarse-graining, which smoothens the finer aspects of residual interactional symmetry. For example, while TYR, TRP and PHE were considered as residues with π-cloud, HIS, ARG, LYS, ASP, GLU – were all considered as charged amino acids, etc.. While calculation of symmetry among such 'classes' of residues has proven helpful for other problems (Banerji and Ghosh, 2011), the coarse-grained description scheme has proved to be inadequate to detect the finer aspects of difference in organization schemes between Ig and non-Ig proteins.

The present work demonstrates how, to ensure Ig stability, nature assigns immense importance to slightest of the differences in symmetry of interaction between any residue i and another residue j. Residue i can be any one of the 20 residues, the same applies for residue j. However, to quantify the finer differences in the extent of symmetry, a generic definition for i-j interactional symmetry would have been inadequate; because in any arbitrarily chosen protein, the extent of space-filling of ALA-GLU interaction symmetry may well be quite different from that quantified for ALA-ASP interaction symmetry, and so on. A generic definition would have considered all residues to be indistinguishable, which, of course, would have proved awfully inadequate to quantify finer differences that characterize Ig and non-Ig structural organization schemes. The present work quantified the symmetry of interaction between distinguishable residues, whereby the coarse-graining induced smoothening of symmetry magnitudes could be avoided. For example, while LYS-ASP attractor (at CD=2.04) is found to be $4^{th}$ deepest Ig attractor, ARG-GLU attractor (at CD=2.26) is found to be the $6^{th}$ deepest Ig attractor. Considering LYS, ASP, ARG, GLU with a coarse-grained class indistinguishable residues (viz. 'charged-residue') would have never enabled us to obtain this information. Owing to the ability to quantify these small differences, residual interaction symmetry could decipher the SAs, which the other symmetry measures could not.

**Acknowledgement:** This work would have never ever taken place without enormous, multidimensional and consistent support of Professor G. Narahari Sastry (CMM, CSIR-IICT, Hyderabad, India). Author is grateful to him.

**Table-1: Ig Attractor Table**

| Ig. Res-Res Pair | Ig. Attractor Magnitude | Ig. CDs residing in attractor | Ig. (Max-Min) of convergent Sequence Length | Ig. Depth of the attractor | Basin of total count (-0.25 space-filling unit) | Basin of total count (+0.25 space-filling unit) | Total number of Ig.s studied for this Res-Res interaction | Basin of normalized count (-0.25 space-filling unit) | Basin of normalized count (+0.25 space-filling unit) | Ig. Total normalized basin count | Ig. Attractor Weightage |
|---|---|---|---|---|---|---|---|---|---|---|---|
| THR-LYS | 2.28 | 3 | 573 | 1719 | 10 | 5 | 29 | 0.345 | 0.172 | 0.517 | 889.138 |
| GLU-LEU | 2.52 | 2 | 932 | 1864 | 8 | 5 | 29 | 0.276 | 0.172 | 0.448 | 835.586 |
| ALA-GLU | 2.36 | 2 | 773 | 1546 | 7 | 7 | 31 | 0.226 | 0.226 | 0.452 | 698.193 |
| LYS-ASP | 2.04 | 2 | 527 | 1054 | 7 | 7 | 26 | 0.269 | 0.269 | 0.538 | 567.538 |
| THR-SER | 2.10 | 2 | 433 | 866 | 8 | 12 | 32 | 0.250 | 0.375 | 0.625 | 541.250 |
| ARG-GLU | 2.26 | 2 | 577 | 1154 | 5 | 6 | 24 | 0.208 | 0.250 | 0.458 | 528.917 |
| PRO-ASN | 2.38 | 2 | 518 | 1036 | 6 | 6 | 28 | 0.214 | 0.214 | 0.428 | 444.000 |
| VAL-ILE | 2.11 | 2 | 512 | 1024 | 6 | 4 | 30 | 0.200 | 0.133 | 0.333 | 341.333 |
| VAL-ILE | 1.63 | 3 | 282 | 846 | 6 | 6 | 30 | 0.200 | 0.200 | 0.400 | 338.400 |
| VAL-ILE | 2.12 | 2 | 442 | 884 | 8 | 3 | 30 | 0.267 | 0.100 | 0.367 | 324.133 |
| ALA-ASP | 1.77 | 3 | 289 | 867 | 3 | 7 | 29 | 0.103 | 0.241 | 0.344 | 298.965 |
| ASN-LEU | 2.00 | 3 | 299 | 897 | 1 | 7 | 28 | 0.038 | 0.250 | 0.288 | 256.286 |
| ASN-PHE | 2.42 | 2 | 364 | 728 | 7 | 2 | 26 | 0.269 | 0.077 | 0.346 | 252.000 |
| ALA-GLN | 2.69 | 2 | 364 | 728 | 4 | 6 | 29 | 0.138 | 0.207 | 0.345 | 251.034 |
| ALA-LYS | 2.40 | 2 | 435 | 870 | 4 | 4 | 28 | 0.143 | 0.143 | 0.286 | 248.571 |
| GLU-ILE | 2.38 | 2 | 364 | 728 | 5 | 3 | 24 | 0.208 | 0.125 | 0.333 | 242.667 |
| LYS-ASP | 2.48 | 2 | 505 | 1010 | 3 | 3 | 26 | 0.115 | 0.115 | 0.230 | 233.077 |
| THR-THR | 2.64 | 2 | 491 | 982 | 6 | 1 | 30 | 0.200 | 0.033 | 0.233 | 229.133 |
| LYS-TYR | 2.36 | 2 | 482 | 964 | 2 | 4 | 26 | 0.077 | 0.154 | 0.231 | 222.461 |
| GLU-GLU | 2.36 | 2 | 476 | 952 | 2 | 3 | 22 | 0.091 | 0.136 | 0.227 | 216.364 |
| ALA-ILE | 1.89 | 3 | 276 | 828 | 3 | 4 | 31 | 0.097 | 0.129 | 0.226 | 186.968 |
| PHE-TYR | 2.39 | 2 | 483 | 966 | 0 | 4 | 21 | 0 | 0.190 | 0.190 | 184.000 |
| ALA-GLU | 2.85 | 2 | 568 | 1136 | 3 | 2 | 31 | 0.097 | 0.064 | 0.161 | 183.226 |
| LEU-LEU | 1.63 | 5 | 114 | 570 | 3 | 5 | 26 | 0.115 | 0.192 | 0.307 | 175.385 |
| ALA-THR | 2.72 | 2 | 547 | 1094 | 4 | 1 | 32 | 0.125 | 0.031 | 0.156 | 170.937 |

**Table-2: Non-Ig Attractor Table**

| Non-Ig. Res-Res Pair | Non-Ig. Attractor Magnitude | Non-Ig. CDs residing in attractor | Non-Ig. (Max-Min) of convergent Sequence Length | Non-Ig. depth of the attractor | Basin of total count (-0.25 space-filling unit) | Basin of total count (+0.25 space-filling unit) | Total no. of Ig.s studied for this Res-Res int. | Basin of normalized count (-0.25 space-filling unit) | Basin of normalized count (+0.25 space-filling unit) | Total normalized basin count | Non-Ig. Attractor Weightage |
|---|---|---|---|---|---|---|---|---|---|---|---|
| ALA-VAL | 2.38 | 2 | 417 | 834 | 5 | 2 | 13 | 0.385 | 0.154 | 0.539 | 449.077 |
| ALA-GLU | 2.23 | 2 | 490 | 980 | 2 | 3 | 13 | 0.154 | 0.231 | 0.385 | 376.923 |
| GLU-ILE | 2.27 | 2 | 306 | 612 | 4 | 2 | 12 | 0.333 | 0.167 | 0.500 | 306.00 |
| ALA-GLN | 2.43 | 2 | 399 | 798 | 1 | 3 | 11 | 0.091 | 0.273 | 0.364 | 290.182 |
| LYS-VAL | 2.48 | 3 | 209 | 627 | 3 | 2 | 11 | 0.273 | 0.182 | 0.455 | 285.00 |
| ALA-GLN | 2.68 | 2 | 384 | 768 | 3 | 1 | 11 | 0.273 | 0.091 | 0.364 | 279.273 |
| GLU-VAL | 2.41 | 2 | 290 | 580 | 3 | 2 | 12 | 0.250 | 0.167 | 0.417 | 241.667 |
| THR-THR | 2.48 | 2 | 388 | 776 | 1 | 2 | 12 | 0.083 | 0.167 | 0.250 | 194.00 |
| ASP-GLY | 2.46 | 2 | 190 | 380 | 3 | 3 | 13 | 0.231 | 0.231 | 0.462 | 175.385 |
| ALA-LYS | 2.50 | 2 | 228 | 456 | 2 | 2 | 13 | 0.154 | 0.154 | 0.308 | 140.308 |
| CYS-ILE | 1.40 | 2 | 209 | 418 | 0 | 2 | 8 | 0 | 0.250 | 0.250 | 104.500 |
| ALA-GLU | 2.29 | 2 | 62 | 124 | 4 | 2 | 13 | 0.308 | 0.154 | 0.462 | 57.231 |
| ASP-ILE | 2.46 | 2 | 83 | 166 | 1 | 3 | 12 | 0.083 | 0.250 | 0.333 | 55.333 |
| TRP-ILE | 2.53 | 2 | 209 | 418 | 1 | 0 | 9 | 0.111 | 0 | 0.111 | 46.444 |
| ALA-THR | 2.61 | 2 | 117 | 234 | 2 | 0 | 13 | 0.154 | 0 | 0.154 | 36.00 |
| ARG-ASP | 1.70 | 2 | 51 | 102 | 2 | 1 | 11 | 0.182 | 0.091 | 0.273 | 27.818 |
| PHE-TYR | 2.00 | 2 | 51 | 102 | 0 | 3 | 13 | 0 | 0.231 | 0.231 | 23.538 |
| LYS-VAL | 1.89 | 2 | 117 | 234 | 0 | 1 | 11 | 0 | 0.091 | 0.091 | 21.273 |
| THR-GLU | 2.19 | 2 | 39 | 78 | 1 | 2 | 13 | 0.077 | 0.154 | 0.231 | 18.000 |
| ARG-ILE | 1.46 | 2 | 51 | 102 | 2 | 0 | 12 | 0.167 | 0 | 0.167 | 17.000 |
| ALA-LYS | 2.20 | 2 | 10 | 20 | 4 | 1 | 13 | 0.308 | 0.077 | 0.385 | 7.692 |
| GLU-ILE | 2.09 | 2 | 11 | 22 | 1 | 3 | 12 | 0.083 | 0.250 | 0.333 | 7.333 |
| ASN-LEU | 2.28 | 2 | 21 | 42 | 0 | 2 | 12 | 0 | 0.167 | 0.167 | 7.000 |
| ALA-ASP | 2.73 | 2 | 19 | 38 | 2 | 0 | 12 | 0.167 | 0 | 0.167 | 6.333 |
| CYS-VAL | 1.57 | 2 | 1 | 2 | 1 | 4 | 10 | 0.100 | 0.400 | 0.500 | 1.000 |

**Figure-1**

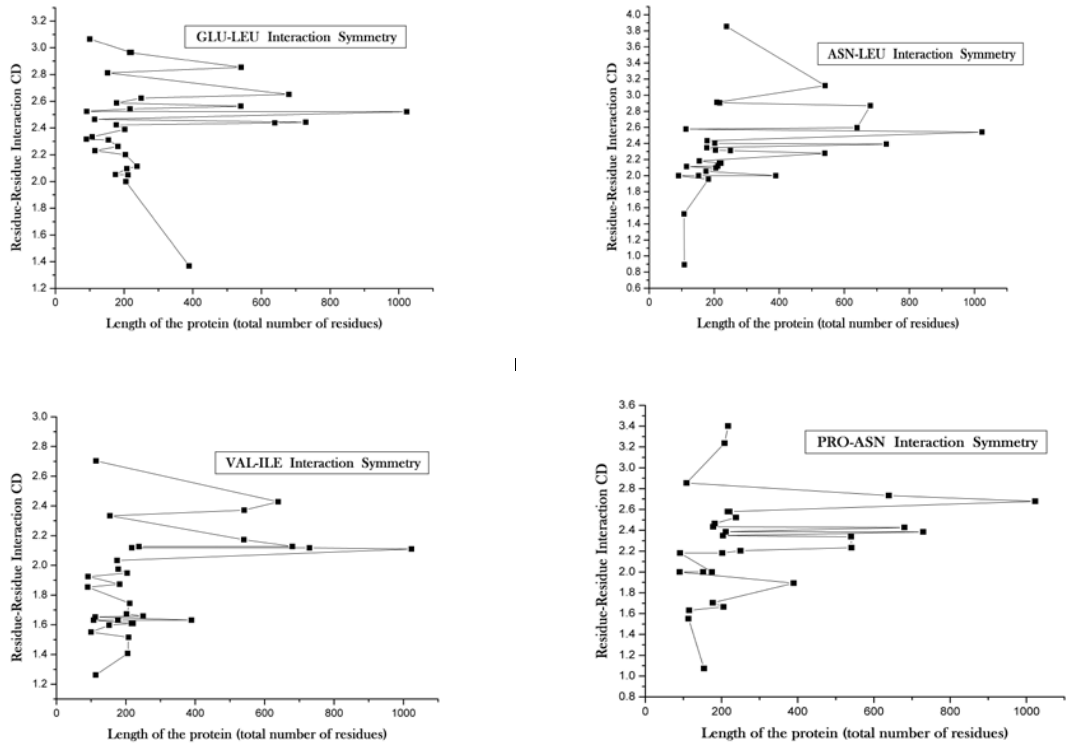

**Figure-1: Structural attractors (of varying types) of Ig proteins**

**Figure-1 and Figure-2 (legend):** For any of these graphs (viz. for pair of residues), the expanse of abscissa (viz. the total number of protein monomers) for any arbitrarily small range of ordinate (viz. for a sufficiently small span of basin of attraction), serves as an easy way to instantly assess the strength of a structural attractor. A strong attractor, due to the innate requirement to maintain the scaffold stability, will invariably be able to make the various extents of residual interaction symmetry to converge onto itself (viz., within an extremely small range of ordinate); notwithstanding the (vastly) disparate total number of residues in these proteins, notwithstanding the difference in the count of the residues (who's interactional symmetry is in question). A small range of ordinate magnitudes with high population of points describes a strong basin for the attractor present therein.

**Figure-2**

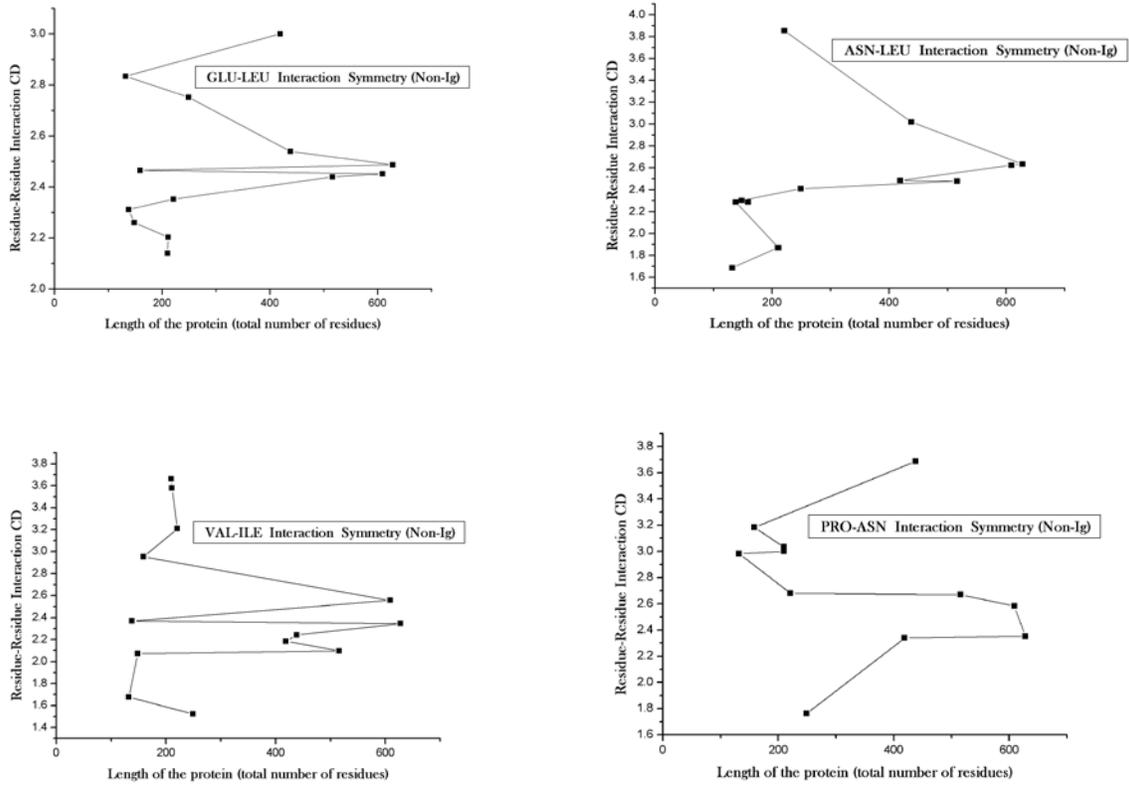

Figure-2: The non-Ig residual interaction symmetries, some of which demonstrate structural attractor like features.

**Figure-3**

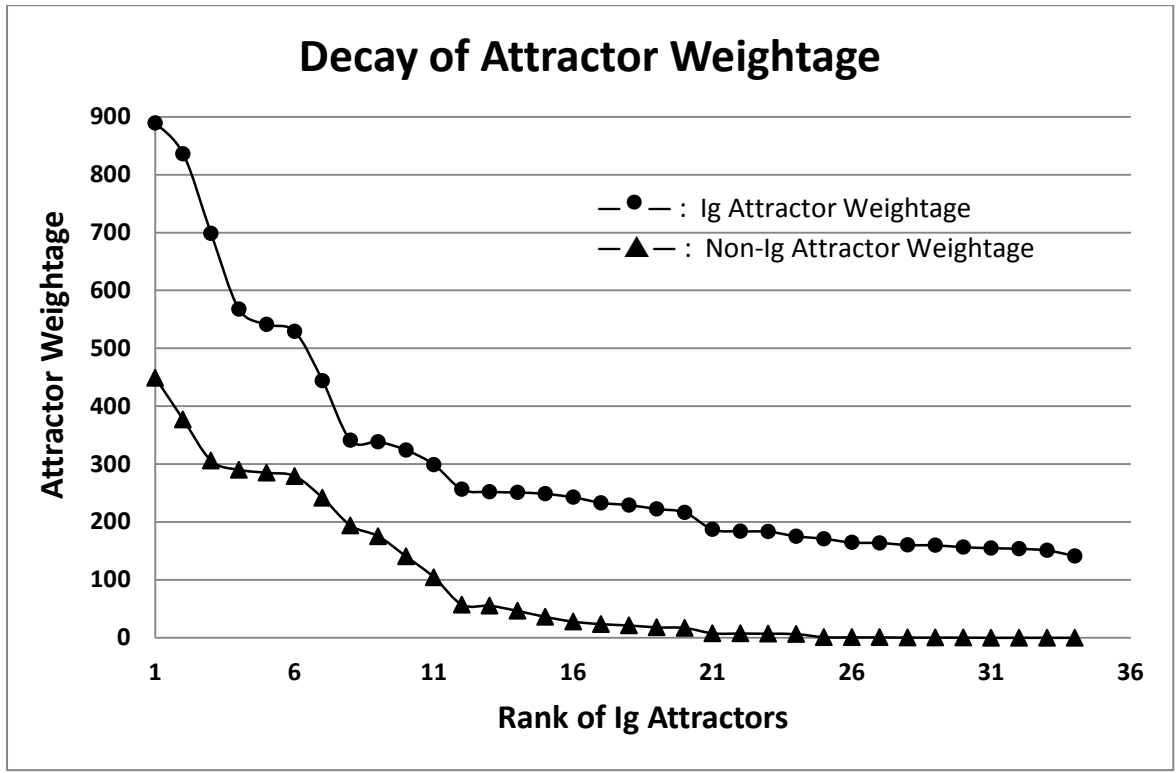

**Figure-4**

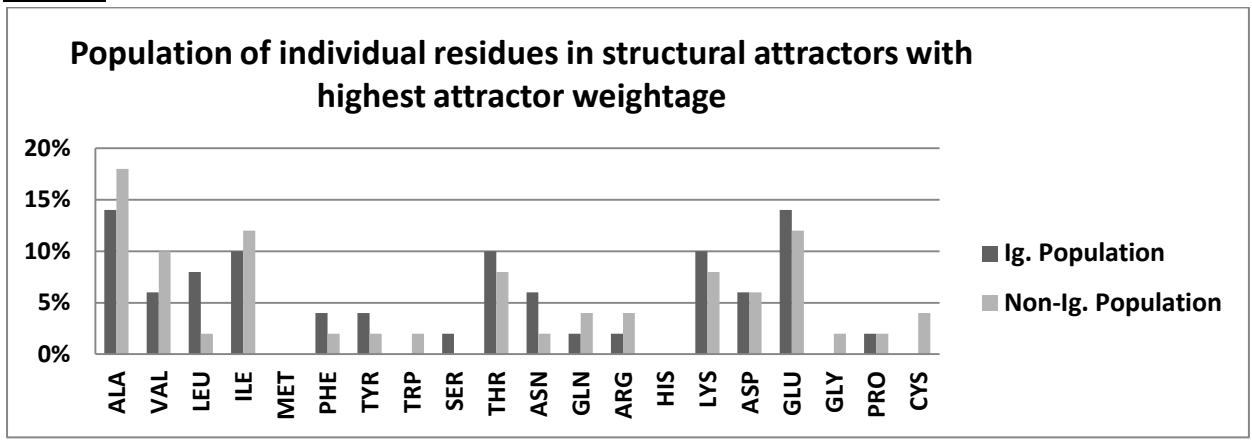

**Figure-5.1: Comparison of symmetry in mass distribution in Ig (Fig.-5.1.B) and non-Ig (Fig.-5.1.A) proteins**

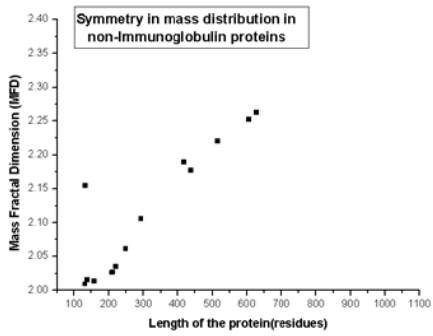
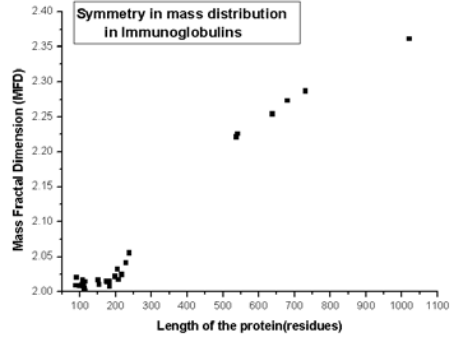

**Figure-5.2: Comparison of symmetry in hydrophobicity distribution in Ig (Fig.-5.2.B) and non-Ig (Fig.-5.2.A) proteins**

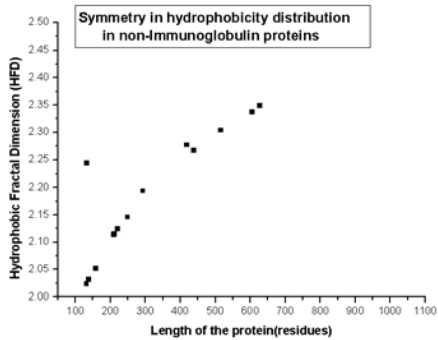
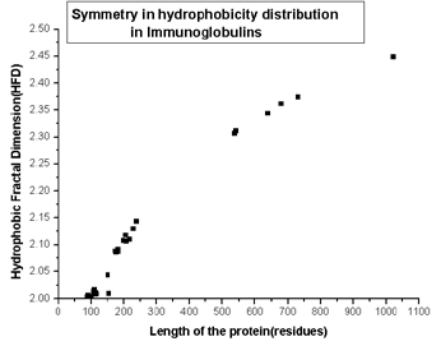

**Figure-5.3: Comparison of unused-hydrophobicity in Ig (Fig.-5.3.B) and non-Ig (Fig.-5.3.A) proteins**

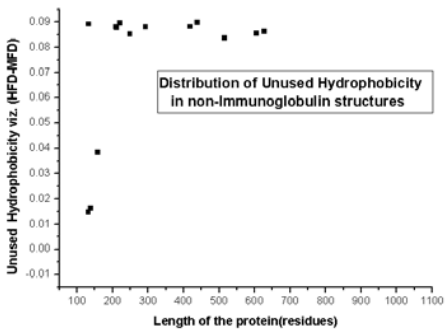
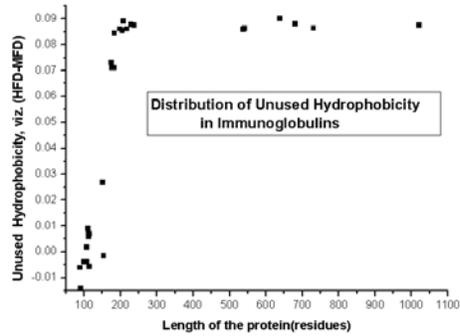

**Figure-5.4: Comparison of symmetry in polarizability distribution in Ig (Fig.-5.4.B) and non-Ig (Fig.-5.4.A) proteins**

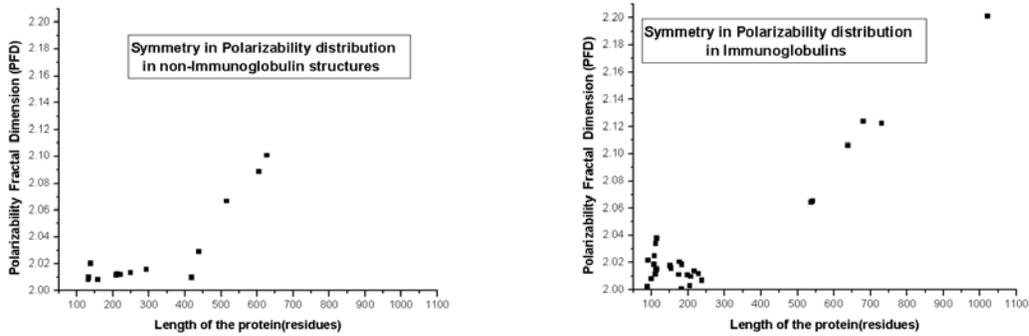

**Figure-5.5: Comparison of symmetry in distribution of peptide-dipole in Ig (Fig.-5.5.B) and non-Ig (Fig.-5.5.A) proteins**

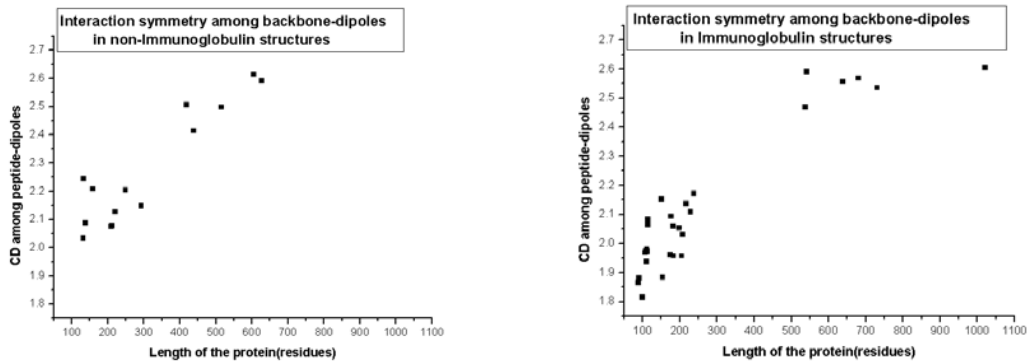

**Figure-5.6: Comparison of symmetry in distribution of charged residues in Ig (Fig.-5.6.B) and non-Ig (Fig.-5.6.A) proteins**

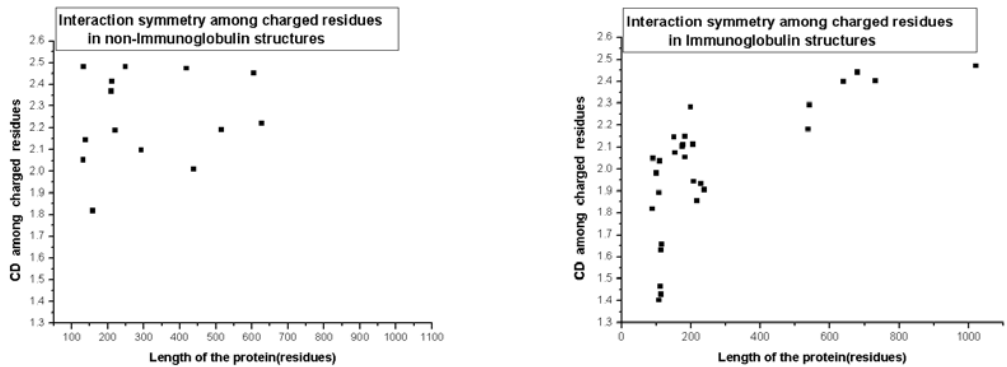

**Figure-5.7:** Comparison of symmetry in distribution of aromatic residues in Ig (Fig.-5.7.B) and non-Ig (Fig.-5.7.A) proteins

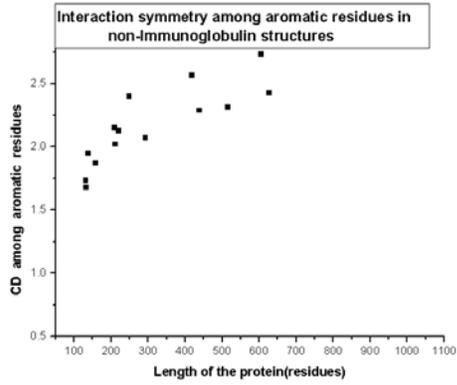
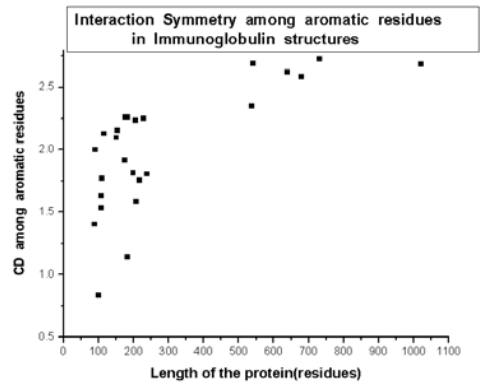

**Figure-5.8:** Comparison of symmetry in distribution of cation-[π]-cloud proximity in Ig (Fig.-5.8.B) and non-Ig (Fig.-5.8.A) proteins

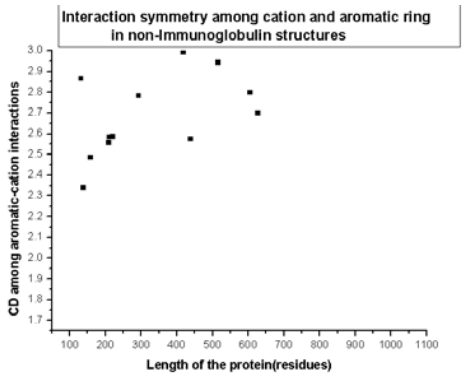
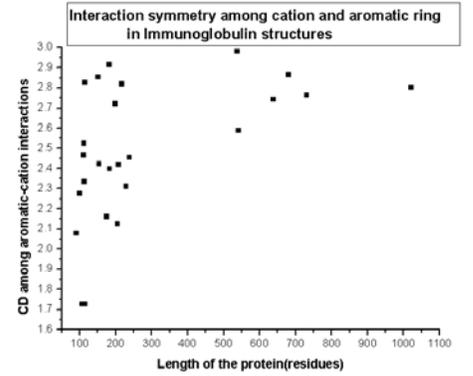

**Figure-5.9:** Comparison of symmetry in distribution of active chiral centers in Ig (Fig.-5.9.B) and non-Ig (Fig.-5.9.A) proteins

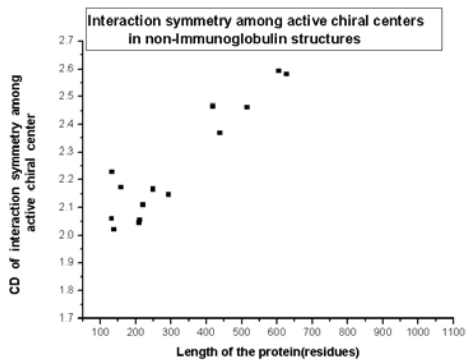
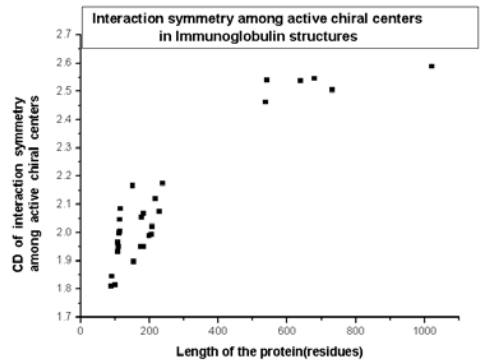

**Figure-5.10: Comparison of symmetry in distribution of hydrophobic residues in Ig (Fig.-5.10.B) and non-Ig (Fig.-5.10.A) proteins**

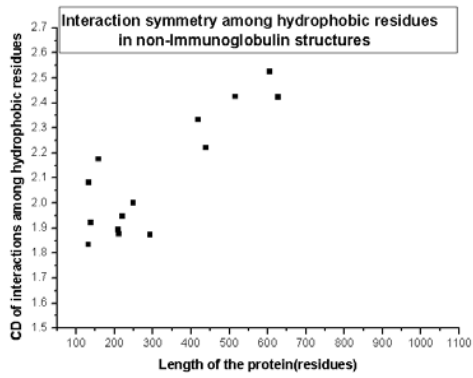 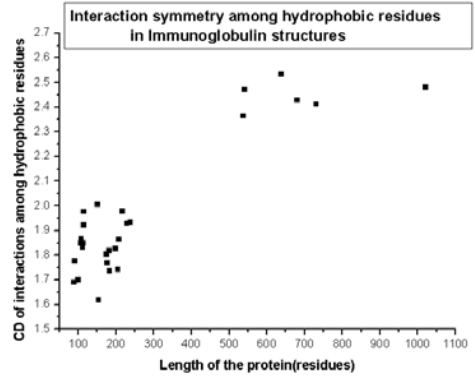